\documentclass[manuscript]{aastex}

\begin{document}

\title{Periodic Variability of the Barium Central Star of the Planetary Nebula
Abell~70}

\author{Howard E.~Bond\altaffilmark{1,2}
and
Robin~Ciardullo\altaffilmark{1}}

\altaffiltext{1}{Department of Astronomy, Pennsylvania State University,
University~Park, PA~16802}

\altaffiltext{2}
{Space Telescope Science Institute, 
Baltimore, MD~21218}

The central stars of planetary nebulae (PNe) are hot objects---the cores of AGB
stars that have recently ejected their outer layers. However, there is a rare
class of ``Abell~35-type'' PN nuclei (PNNi), with optical spectra dominated by
cool stars (Bond et~al.\ 1993). These are likely binary companions of the
visually fainter hot cores. In the first three known A~35-type nuclei (A~35
itself, LoTr~1, and LoTr~5), a rapidly rotating late-type giant or subgiant is
seen optically, while the hot PNN is detected at UV wavelengths. 

There are several known wide binaries containing hot white dwarfs and cool,
rapidly rotating companions, which are probably descendants of A~35-type PNNi.
Boffin \& Jorissen (1988) and Jeffries \& Stevens (1996) proposed that these
systems result from an AGB star in a wide binary developing a dense stellar
wind, part of which accretes onto a late-type companion, spinning it up to a
faster rotation. 

If the outer layers of the AGB star contain nuclearly processed material
(carbon, {\it s}-process elements), the surface of the wind-accreting companion
will become contaminated. Such a process appears to account for the origin of
late-type ``barium stars,'' first identified by Bidelman \& Keenan (1951), and
later shown to have a high fraction of unseen binary companions (McClure 1984).

Support for this scenario came from the discovery (Bond et~al.\ 2003) that the
PNN of WeBo~1 is a cool barium star, with enhanced carbon and {\it s}-process
elements. UV photometry by the {\it Neil Gehrels Swift Observatory\/} confirmed
that WeBo~1 has a hot companion (Siegel et~al.\ 2012). The cool component is
chromospherically active, with an apparent rotation period of 4.7~days, based on
periodic photometric variability and the likely presence of starspots. 

Recently, several more cool Ba stars have been found in A~35 PNe (e.g.,
Miszalski et~al.\ 2013, Tyndall et~al.\ 2013, Jones \& Boffin 2017). In many
cases the surrounding PNe have a pronounced ring-like morphology.  In addition
to WeBo~1, rotation periods of a few days have been found for the cool
components of LoTr~1 (5.95~d; Aller et~al.\ 2018), LoTr~5 (6.4~d; Tyndall
et~al.\ 2013), and Hen~2-39 ($\sim$5.5~d; Miszalski et~al.\ 2013). The binary
orbital periods, when known, are much longer, e.g., $\sim$2700~d for LoTr~5.

The central star of the ``diamond-ring'' PN Abell~70 (hereafter A~70) was found
to be a cool barium-rich star by Miszalski et~al.\ (2012, hereafter M12). Its
optical spectrum shows a G8~IV-V subgiant, but UV data confirm the presence of a
hot PNN in the binary. In order to investigate whether the Ba star in A~70 
exhibits a short rotation period, we obtained optical photometry during 2010 and
2011, and report the results here.

Observations were made by Chilean service personnel, using the 1.3-m SMARTS
Consortium\footnote{SMARTS is the Small \& Moderate Aperture Research Telescope
System; {\tt http://www.astro.yale.edu/smarts}} telescope at CTIO and its
ANDICAM CCD camera. Data were obtained on 25 nights between 2010 October~26 and
November~28 ($I$~band only), and 79 nights between 2011 July~30 and November~29
($B$, $V$, and $I$ bands). The frames were bias-subtracted and flat-fielded in
the SMARTS pipeline at Yale University. We then carried out photometry, using
standard IRAF tasks, as described in Bond et~al.\ (2016), to determine relative
magnitudes of A~70 and four nearby comparison stars. We then calculated the
magnitude difference between A~70 and the sum of the fluxes of the comparison
stars. On two photometric nights, we determined approximate calibrated
magnitudes for A~70 by reference to a standard field of Landolt (1992),
obtaining $B=18.5$, $V=17.8$, and $I=16.9$, in good agreement with M12. The
recent {\it Gaia\/} DR2 parallax (Gaia Collaboration et~al.\ 2018) of
$0.3829\pm0.1657$~mas gives an approximate absolute magnitude of
$M_V\simeq+5.7$, consistent with a late G star near the main sequence.

Our 2010 data clearly show a 2.061~d, periodic variation, as shown in the phased
$I$-band magnitudes at the top of Figure~1.  Because this period is so close to
2~days, at least one alias period, 1.939~days, fits the data nearly as well.
However, when the observations resumed in 2011, this variation had become
undetectable, as shown in the $BVI$ light curves labeled ``2011a'' in Figure~1.
By about 2011 August~20, the variations reappeared, at the same period but with
lower amplitude, as shown in the curves labelled ``2011b.'' This behavior is
consistent with an origin in starspots, whose amount of coverage varies with
time as the activity level changes. (The rotation period could, of course, be
twice the value given, if there are, for example, two starspots.)

A~70 thus joins other A~35-type PNNi in having a rotating, late-type spotted
companion of the hot central star. It would be of interest to determine the
orbital period of the system, based on long-term radial-velocity monitoring.

\acknowledgments

H.E.B. thanks the STScI Director's Research Fund for supporting participation in
the SMARTS consortium. The CTIO/SMARTS service observers were Juan~Espinoza,
Rodrigo~Hernandez, Alberto~Miranda, and Jacqueline~Seron.

\begin{figure}
\begin{center}
\includegraphics[height=6in]{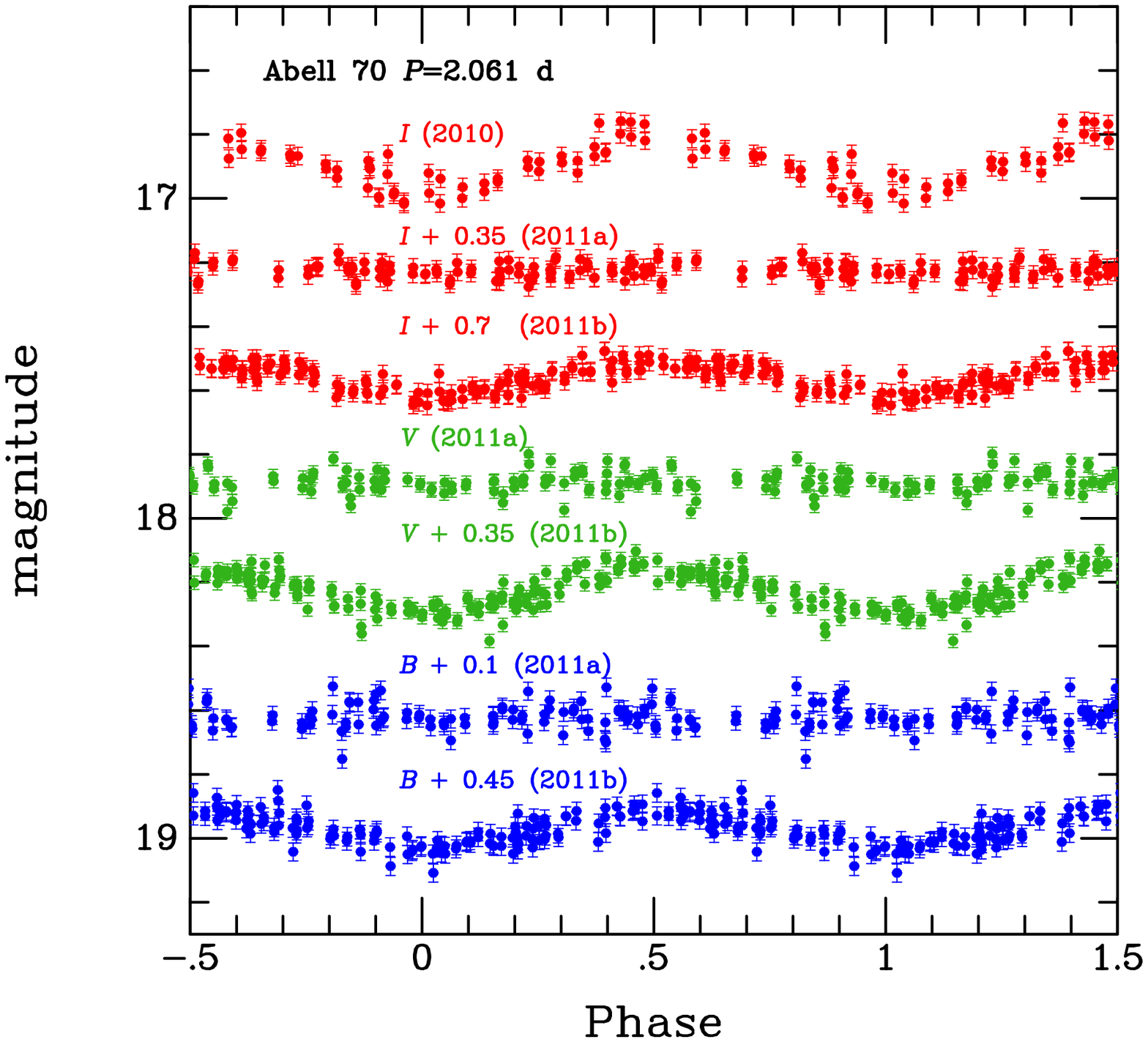}
\figcaption{\footnotesize
Phased $BVI$ light curves of A~70 obtained in 2010 ($I$-band only),
2011.58--2011.70 (``2011a''), and 2011.70--2011.91 (``2011b''). Zero-points are
offset by the amounts indicated in the figure. The 2.061~d variation was present
in 2010 and 2011b, but absent during 2011a.
}
\end{center}
\end{figure}

\end{document}